\documentclass[aps, prl,showpacs,twocolumn, superscriptaddress]{revtex4}

\usepackage{graphicx}
\usepackage{dcolumn}
\usepackage{bm}
\usepackage{color}
\usepackage{amsmath}
\usepackage{amssymb}

\begin{document}

\title{Generation and distribution of high-dimensional frequency-entangled qudits}

\author{Rui-Bo Jin}
\email{jrbqyj@gmail.com}
\affiliation{National Institute of Information and Communications Technology (NICT), 4-2-1 Nukui-Kitamachi, Koganei, Tokyo 184-8795, Japan}
\author{Ryosuke Shimizu}
\affiliation{University of Electro-Communications (UEC), 1-5-1 Chofugaoka, Chofu, Tokyo 182-8585, Japan}
\author{Mikio Fujiwara}
\affiliation{National Institute of Information and Communications Technology (NICT), 4-2-1 Nukui-Kitamachi, Koganei, Tokyo 184-8795, Japan}
\author{Masahiro Takeoka}
\affiliation{National Institute of Information and Communications Technology (NICT), 4-2-1 Nukui-Kitamachi, Koganei, Tokyo 184-8795, Japan}
\author{Ryota Wakabayashi}
\affiliation{National Institute of Information and Communications Technology (NICT), 4-2-1 Nukui-Kitamachi, Koganei, Tokyo 184-8795, Japan}
\affiliation{Waseda University, 3-4-1 Okubo, Shinjyuku, Tokyo 165-8555, Japan}
\author{Taro Yamashita}
\affiliation{National Institute of Information and Communications Technology (NICT), 588-2 Iwaoka, Kobe 651-2492, Japan}
\author{Shigehito Miki}
\affiliation{National Institute of Information and Communications Technology (NICT), 588-2 Iwaoka, Kobe 651-2492, Japan}
\author{Hirotaka Terai}
\affiliation{National Institute of Information and Communications Technology (NICT), 588-2 Iwaoka, Kobe 651-2492, Japan}
\author{Thomas Gerrits}
\affiliation{National Institute of Standards and Technology  (NIST), 325 Broadway, Boulder, Colorado 80305, USA}
\author{Masahide Sasaki}
\email{psasaki@nict.go.jp}
\affiliation{National Institute of Information and Communications Technology (NICT), 4-2-1 Nukui-Kitamachi, Koganei, Tokyo 184-8795, Japan}

\date{\today }

\begin{abstract}
We demonstrate a novel scheme to generate frequency-entangled qudits with dimension number higher than 10 and to distribute them over optical fibers of 15\,km in total length.
This scheme combines the technique of spectral engineering of biphotons generated by spontaneous parametric down-conversion and the technique of spectrally resolved  Hong-Ou-Mandel interference.
We characterized the comb-like spectral correlation structures of the qudits by time of arrival measurement and correlated spectral intensity measurement.
The generation and distribution of the distinct entangled frequency modes may  be useful for quantum cryptography, quantum metrology, quantum remote synchronization, as well as fundamental test of stronger violation of local realism.

\end{abstract}



\maketitle

\textbf{\emph{Introduction}}
The study of higher dimensional entangled states has opened a wide range of possibilities in quantum information science and technology.
For example, $d$-dimensional quantum bits (qudits) in principle have a stronger violation of the local hidden variable hypothesis than
a qubits system \cite{Kaszlikowski2000}, and are even useful to observe the violation of its more general context including non-contextuality \cite{Lapkiewicz2011}.
Entangled qudits are also advantageous in quantum key distribution (QKD)\cite{Durt2004, Nunn2013, Mower2013}, and other applications such as quantum bit commitment \cite{Langford2004} or spatial imaging \cite{Howland2013}.

Several experiments have been dedicated to prepare photonic qudit states with photons encoded in different degrees of freedom.
For example, qudits  have been generated from biphotons in  polarization \cite{Bogdanov2004}, frequency \cite{Bernhard2013}, time-bin \cite{Thew2004, Bessire2014}, spatial modes \cite{Krenn2014}.
Among all these  degrees of freedom, frequency may be of great interest in distributing entangled photons in multiplexed modes  over long-distance optical fibers for quantum communication. Traditionally,  the frequency-entangled qudits can be generated through manipulating the spectra of biphotons by using narrow-band filters and electro-optic modulators \cite{Olislager2010}, or by using a spatial light modulator (SLM) \cite{Bernhard2013}.
However, the separation and measurement of these high-dimensional frequency-entangled state is limited by optical loss and resolution of the devices.
Therefore, up to now, only 5-dimensional qudits ($d$=5) were experimentally generated \cite{Olislager2010}.

In this Letter, we experimentally demonstrate a novel method to generate qudits with $d$ higher than 10 and to distribute the state over 15\,km distance.
Our scheme is based on two key techniques.
The first one is the quantum state engineering of twin photons from type-II spontaneous parametric down-conversion (SPDC).
We employ a periodically poled MgO-doped stoichiometric LiTaO$_3$ (PPSLT) crystal \cite{Shimizu2009, Bisht2015}, which satisfies the group-velocity matching (GVM) condition of  $V^{-1}_{s}=V^{-1}_{i}$ at telecom wavelength, where $V^{-1}_{s(i)} $ is the inverse of the group velocity  for  the signal (idler) photons from SPDC \cite{Edamatsu2011}. This GVM condition allows us to generate a broadband
twin photons with strong spectral anti-correlation.
The second technique is the spectrally resolved Hong-Ou-Mandel (HOM) interference \cite{Hong1987}, where the spectral modes of the twin photons can be separated at different delay positions in the interference \cite{Gerrits2015, Jin2015OE}.
By combining these two techniques, a multi-peak (comb-like) spectral correlation structure can be formed in a two-dimensional diagram of the twin photon frequencies which is observed by coincidence counts at the two distant nodes after optical fiber transmission.

\begin{figure}[tbp]
\centering
\includegraphics[width= 0.48\textwidth]{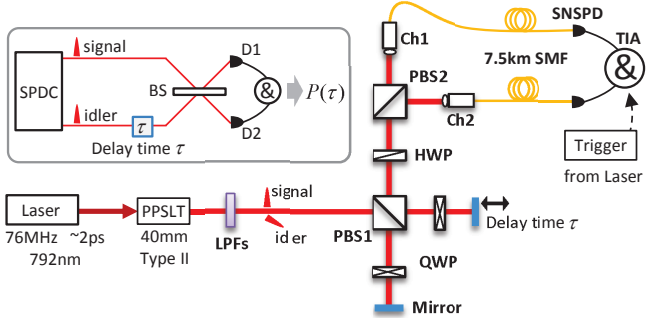}
\caption{ (color online) The experimental setup.
Laser pulses (76\,MHz, 792\,nm, 50\,mW  average power, temporal duration $\sim$ 2\,ps) from a mode-locked Titanium sapphire laser 
are focused into a 40-mm-long PPSLT crystal with a polling period of 21.5\,$\mu$m for a type-II collinear SPDC.
The temperature is maintained at $40.8\,^{\circ}\mathrm{C}$ to achieve degenerate wavelengths for the signal and idler photons, which are polarized horizontally (H) and vertically (V), respectively.
After filtered by long pass filters (LPFs), the downconverted photons with central wavelength of 1584\,nm and FWHM (full width at half maximum) of 22\,nm pass through a polarization beam splitter (PBS1), a half wave plate (HWP, at 22.5\,$^{\circ}$) and PBS2 for the HOM interference.
Two quarter wave plates (QWP, at 45\,$^{\circ}$) placed before two mirrors are used to convert the H(V)-polarized photon to be V(H)-polarized after a round pass.
Finally the photons are coupled into two 7.5-km-long commercial available single-mode fibers (SMF), 
 and detected by two superconducting nanowire single photon detectors (SNSPDs), which are connected to a Time Interval Analyzer (TIA).
The electrical trigger signals from the laser provide timing information of the laser pulses to TIA for synchronization in the correlated spectral intensity (CSI) measurement.
The inset depicts a standard configuration of the HOM interference, which is a simplified model of the experimental setup.
}
\label{setup}
\end{figure}

\textbf{\emph{Schematic}}
The schematic of our experiment is shown in Fig.\,\ref{setup}.
%
%
%
%
%
%
It is a simple HOM interference setting with an additional time delay $\tau$
in the idler mode.
Quantum state of the photon pair emitted from a SPDC source
is \cite{URen2006, Jin2013OE}
\begin{equation}\label{SPDC}
|\Psi\rangle = \int^\infty_0 \int^\infty_0 d\omega_1 d\omega_2
f(\omega_1,\omega_2)
\hat{a}_s^\dagger (\omega_s) \hat{a}_i^\dagger (\omega_i) |0\rangle ,
\end{equation}
where $\hat{a}^\dagger(\omega)$ is the creation operator at angular frequency
$\omega$, the subscripts $s$ and $i$ denote the signal and idler photons, respectively, and $f(\omega_s, \omega_i)$ is their joint spectral amplitude (JSA).

After applying the delay and the HOM beamsplitter, and
postselecting the state such that each output channel
(Ch1 and Ch2 in Fig.\,\ref{setup}) contains one photon,
and taking into account the structure of $f(\omega_s, \omega_i)$
for our crystal and pumping,
the state is approximately described as
\begin{eqnarray}\label{EntQudit}
|\phi (\tau)\rangle & \approx & \frac{1}{\sqrt{\mathcal{N}}}
\int^\infty_0 \int^\infty_0 d\omega_1 d\omega_2
h(\omega_1 + \omega_2 - \omega_p)
\nonumber\\ &&
\times \delta (\omega_1 + \omega_2 - \omega_p)
(1-e^{-i(\omega_1-\omega_2)\tau})
\nonumber\\ &&
\times \hat{a}_1^\dagger (\omega_1) \hat{a}_2^\dagger (\omega_2) |0\rangle ,
\end{eqnarray}
where $\mathcal{N}$ is a normalization factor,
$\omega_p$ is the pump frequency, the subscripts $1$ and $2$ denote modes in Ch1 and Ch2,
$\delta(x)$ is the Dirac delta function, and
$h(x)$ is a function determined by the phase matching condition
and the pump envelope.
In Eq.~(\ref{EntQudit}), one can observe that
the amplitude of the photon-pair wave function distributes only on
$\omega_1 + \omega_2 = \omega_p$ and oscillates
with peaks at $\omega_1-\omega_2=2\pi /\tau$, that is,
having an entangled qudit structure
$\sim \sum_j^m c^j |\omega_+^j \omega_-^j\rangle$
where $\omega_\pm^j = \omega_p/2 \pm (2j-1)\pi /\tau$.
See Supplementary Information 1 for more detailed discussion.

\textbf{\emph{Experiment}}
To generate such a state, one needs 1) high indistinguishability between the signal and idler photons from the SPDC source, and 2) a carefully engineered JSA showing discrete peaks in frequency correlation.
These properties are experimentally characterized as follows.

\begin{figure}[tbp]
\centering
\includegraphics[width= 0.3 \textwidth]{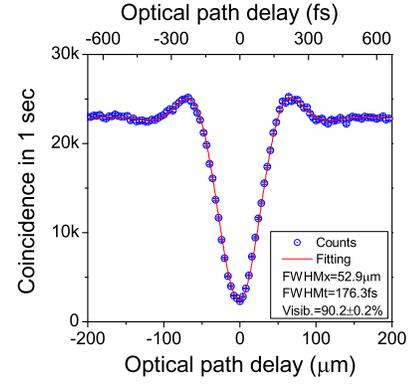}
\caption{ (color online) The Hong-Ou-Mandel dip. Coincidence counts in one second as a function of the optical path delay. The error bars were evaluated by assuming Poissonian statistics of these coincidence counts.}
\label{dip}
\end{figure}

First, the indistinguishability was observed by measuring
the coincidence probability $ P(\tau )$ in the HOM interference
(Fig.\,\ref{setup}, inset) as a function of the delay time $\tau$ \cite{Grice1997}, where
\begin{equation}\label{P(tau)}
P(\tau)= \frac{1}{4}\int_0^\infty  {\int_0^\infty  d } \omega _1 d\omega _2 I(\omega _1 ,\omega _2 ,\tau )
\end{equation}
where
\begin{equation}\label{JSI}
I(\omega _1 ,\omega _2 ,\tau ) =
|f(\omega_1, \omega_2) - f(\omega_2, \omega_1)
e^{-i(\omega_1-\omega_2)\tau}|^2 ,
\end{equation}
is the correlated spectral intensity
(CSI) (see Supplementary Information 2 for the detailed calculation).
The measured HOM dip is shown in Fig.\,\ref{dip}.
The net visibility is evaluated as 90.2$\pm$0.2\%, indicating high indistinguishability between the signal and idler photons from our SPDC source.
The FWHM of the dip is 52.9\,$\mu$m (176.3\,fs).
The raw visibility without subtracting the background counts  is 68.6$\pm$0.4\%.
%
%
%
%
%
%
%
%
%

Next, we characterize the comb-like structure of frequency-entangled qudits using one-dimensional time-of-arrival (ToA) measurement and two-dimensional CSI measurement.
%
%
\begin{figure*}[tbp]
\centering
\includegraphics[width= 0.98 \textwidth]{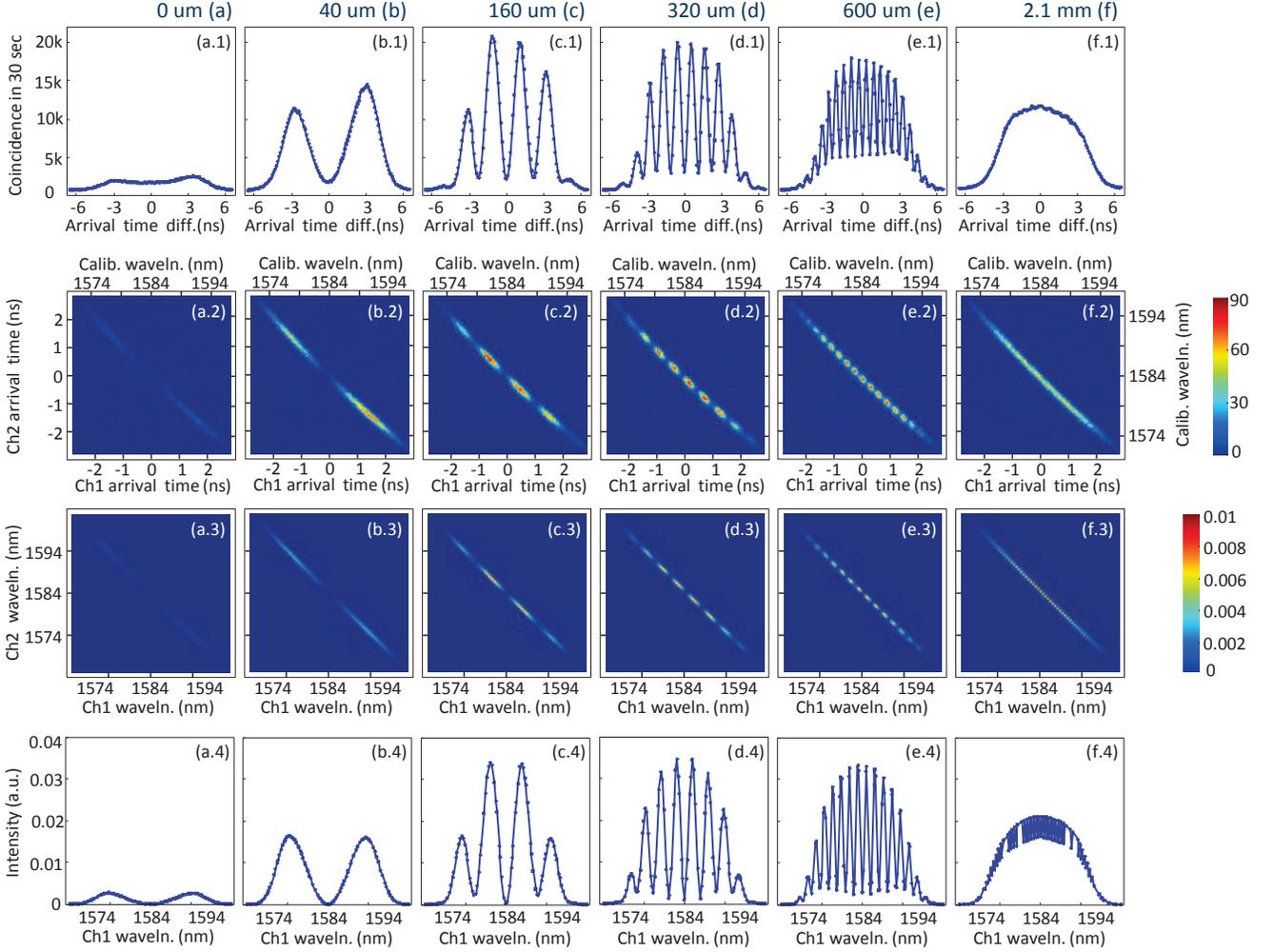}
\caption{ (color online) Experimental comb-like structures with theoretical simulations.
The first row (a.1-f.1) shows experimental ToA data $H(\Delta \omega,\tau)$. The horizontal axis is the arrival time difference $t_2 - t_1$, corresponding to $\Delta \omega=\omega_2-\omega_1$.
The second row (a.2-f.2) shows experimental correlated spectral intensity (CSI) $I(\omega_1,\omega_2,\tau)$.
Graphs (a-f) are at the delay positions of 0\,$\mu$m, 40\,$\mu$m, 160\,$\mu$m, 320\,$\mu$m, 600\,$\mu$m and 2.1\,mm, correspoding to the delay time $\tau=$  0\,ps, 0.13\,ps, 0.53\,ps, 1.07\,ps, 2.0\,ps and 7.03\,ps in the HOM dip.
Theoretically simulated data are presented in the third row (a.3-f.3) for the CSI and in the forth row (a.4-f.4) for the marginal spectra at the same delay positions. }
 \label{comb}
\end{figure*}
%
%
In both measurement, we employ a fiber-based spectrometer \cite{Avenhaus2009, Gerrits2011}, which is constituted of  two 7.5-km-long single-mode fibers (SMFs), two superconducting nanowire single photon detectors (SNSPDs) and a  time interval analyzer (TIA), as shown in Fig.\,\ref{setup}.
The wavelength dispersion effect in optical fibers is used to convert spectral information to  arrival time information.
We demonstrated clear comb-like structures, not only in the two-dimensional $(\omega_1, \omega_2)$ diagram, but also in the one dimensional ToA data $H(\Delta \omega,\tau)$,  after integrating the frequency of Ch1 ($\omega_1$):
\begin{equation}\label{eq4}
H(\Delta \omega,\tau)= \int_0^\infty d\omega_1 I(\omega_1,\omega_1+\Delta\omega,\tau),
\end{equation}
where $\Delta \omega = \omega_2-\omega_1$.
In the ToA measurement, the TIA provides the information of the frequency difference  $\Delta \omega=\omega_2-\omega_1$.
Figure\,\ref{comb} shows these comb-like structures, corresponding to frequency-entangled qudits.
Graphs in the first row (a.1-f.1) show the experimental results of the ToA data $H(\Delta \omega,\tau)$. The horizontal axis is the arrival time difference $t_2 - t_1$, corresponding to frequency difference $\Delta \omega=\omega_2-\omega_1$.
The ToA data were accumulated for 30\,seconds with Ch1 as the stop channel and Ch2 as the start channel for the  TIA, at a fixed delay time  $\tau$.
At the dip center, i.e. zero delay position, there is no comb-like structure, as shown in Fig.\,\ref{comb}(a.1).
As the delay position is moved from zero to 40\,$\mu$m, 160\,$\mu$m, 320\,$\mu$m, and 600\,$\mu$m, the peak numbers increased to 2, 4, 8 and 14, as shown in Fig.\,\ref{comb}(b.1-e.1).
When the delay position is  far away from the dip center, the comb-like structure disappears gradually.  %
Fig.\,\ref{comb}(f.1) shows such a case at the delay position of 2.1\,mm.

Graphs in the second row (a.2-f.2) show experimental CSI $I(\omega_1,\omega_2,\tau)$ of the twin photons.
The CSI data are measured with three input signals to the TIA: the trigger (generated by a fast photodiode, recording the timing information of the pump laser pulses for synchronization) as the start channel,  Ch1 as the first stop channel, and  Ch2 as the second stop channel.
The CSI is reconstructed by analyzing the arrival time of the photon pairs in Ch1 and Ch2.
Each CSI dataset was accumulated only for 5\,seconds, thanks to our highly efficient fiber spectrometer which is constitute of high-efficiency SNSPDs \cite{Miki2013, Yamashita2013, Jin2015OC}.
The peak numbers in the ToA data in the first row of Fig.\,\ref{comb} correspond well to the discrete mode numbers  in the diagram $(\omega_1, \omega_2)$ of the CSI in the second row of Fig.\,\ref{comb}.
The mode numbers in Fig.\,\ref{comb}(b.2-e.2) are 2, 4,  8 and 14, respectively.
In  Fig.\,\ref{comb}(f.2), the correlation diagram does not show any mode structure due to our limited timing resolution.
In the CSI data, the comb position is determined by $\mathrm{cos}[(\omega _1  - \omega _2 )\tau]=-1$ in Eq.(\ref{JSI}).
The peak-to-peak distance is equal to $\omega _1  - \omega _2 = 2 \pi / \tau $.
It can be noticed that the full width of the ToA data ($\sim$8\,ns in Fig.\,\ref{comb}(f.1)) is about two times of the full width of the CSI data ($\sim$4\,ns in Fig.\,\ref{comb}(f.2)).
This difference is caused by the fact that $\omega _1 (\omega _2)$ has an arrival time range of around [-2 ns, 2 ns], while the difference of $\omega _1 $ and   $\omega _2 $, i.e. $\omega _1 - \omega _2 $, has an arrival time range of around [-4 ns, 4 ns].

We also performed numerical simulation of the CSI using the experimental conditions  (40-mm-long PPSLT crystal; 2-ps-long pump pulse with Gaussian shape; filtering by a LPF with experimentally measured transmission spectrum), as shown in Fig.\,\ref{comb}(a.3-f.3).
The experimentally measured frequency mode numbers coincide well with our simulation.
The experimental sizes of the comb teeth in Fig.\,\ref{comb}(a.2-f.2) are slightly ``fatter'' than the theoretical sizes in Fig.\,\ref{comb}(a.3-f.3), due to the limited timing resolution of the detection system.
By integrating the data in Fig.\,\ref{comb}(a.3-f.3) over Ch2 wavelength, we obtain the marginal spectra of Fig.\,\ref{comb}(a.3-f.3) over  Ch1 wavelength, as shown in Fig.\,\ref{comb}(a.4-f.4).
The peak numbers in Fig.\,\ref{comb}(a.3-f.3) match well to the peak numbers in Fig.\,\ref{comb}(a.1-f.1).
The profile of Fig.\,\ref{comb}(f.4) also agrees well with that of Fig.\,\ref{comb}(f.1).

\begin{figure}[tbp]
\centering
\includegraphics[width= 0.4 \textwidth]{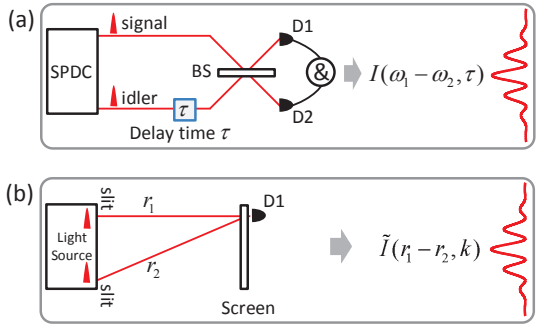}
\caption{ (color online)
The phenomenon of the comb-like structure in the the HOM interference (a) can be intuitively compared with the Young's double slit interference (b).
} \label{model}
\end{figure}

\textbf{\emph{Discussion}}
To intuitively understand the phenomenon of comb-like structure of frequency-entangled qudits in this work, we compare the HOM interference with the classical Young's double slit interference, as shown in Fig.\,\ref{model}(a, b).
%
%
%
Eq.\,(\ref{JSI}) can be rewritten as:
\begin{equation}\label{analogy1}
I(\omega _1  - \omega _2 ,\tau ) = I_1  + I_2  - 2\sqrt {I_1 I_2 } \mathrm{cos}(\omega _1  - \omega _2 )\tau,
\end{equation}
where $I_1={\rm{|}}f(\omega _1 ,\omega _2 ){\rm{|}}^{\rm{2}} $ and $ I_2 = {\rm{|}}f(\omega _2 ,\omega _1 ){\rm{|}}^{\rm{2}}$.
In the classical Young's double slit interference,  as shown in Fig.\,\ref{model}(b), the fringes can be described by
\begin{equation}\label{analogy2}
\tilde I(r_1  - r_2 ,k) = \tilde I_1  + \tilde I_2  - 2\sqrt {\tilde I_1 \tilde I_2 } \cos (r_1  - r_2 )k,
\end{equation}
where $\tilde I$ is the overall intensity; $\tilde I_1$ and $\tilde I_2$ are the intensity of light from the double slit; $k$ is the wave number of the light.
Eq.\,(\ref{analogy1}) and Eq.\,(\ref{analogy2}) have the same mathematical structure.
The physics of HOM interference and double slit interference are different: the former is a two-photon effect which reflects the indistinguishability of photons, while the latter is the light wave interference in the classical sense and also the interference of a single photon itself in the quantum domain.  However, the manipulation of comb teeth structure in the interference pattern can be done in the similar way in which one can find clear correspondence between the parameters in the two schemes: by shifting the frequency difference $\omega _1  - \omega _2$ (optical path difference of $r_1 - r_2$), we can adjust the dark or bright points of the interference pattern; by changing the delay time $\tau$ (wave vector $k$), we can vary the teeth numbers in the comb-like structure.
Thus, the mechanism of our comb-like structure can be intuitively explained along with the simple and fundamental phenomenon in optics.

\textbf{\emph{Conclusion}}
We have proposed and demonstrated the generation of frequency-entangled qudits in the spectrally resolved HOM interference using the downconverted photons from a PPSLT crystal.
The observed maximal spectral mode number is 14 (Fig.\,\ref{comb}) after transmitting 15\,km.
It is worth to note that the number can be improved simply by removing the LPFs in Fig.\,\ref{setup}. By doing this, we are able to observe
more than 20 spectral modes where the FWHM of the downconverted photons is measured to be $\sim$35 nm and the net visibility of the HOM dip is around 75\%.

There are several interesting future issues.
First, the dimensionality of entangled qudits could be further improved by using more broadband twin photon sources \cite{Tanaka2012}.
Another important challenge is more rigorous characterization of entanglement for the generated frequency-entangled qudit states.
To rigorously ``prove'' entanglement without modeling (assuming) physical system of the setup, one has to experimentally access some entanglement measure or
separability criteria which requires state tomography or similar kind of detection.
Such experimental techniques have been well established for standard entangled qubit photons, such as polarization entanglement \cite{James2001}.
However, its application to our high-dimensional frequency-entangled qudits is not straightforward since it is a higher dimensional system and
more importantly, the measurement basis should include the superposition of many different frequency modes which is technically nontrivial to implement.
Recent progress on frequency mixed detection \cite{Bernhard2013} might be a promising way if it is extendable to higher dimensional system including more than 10 frequency modes.

Also, our techniques developed here have various potential applications.
For example, the spectrally resolved HOM interference technique used in this experiment would be useful to extend the dynamic range of the phase shift measurement and timing synchronization, because this technique allows one to see the interference patterns for a longer delay position or delay time than that in the standard HOM interference.
Actually, even when the twin photons have no interference pattern in the region with the delay position larger than 160 $\mu$m in Fig.\,\ref{dip} for the standard time-domain HOM interference, we can still see the interference pattern in frequency domain, as shown in Fig.\,\ref{comb}(c.1-e.1). 
This property may be used for quantum remote synchronization \cite{Giovannetti2001,Ho2009, Quan2016}.  The comb-like structure may also be useful in metrology \cite{Ren2012, Ren2014}.

Finally, our frequency entangled source is clearly applicable to the recent proposals of the high-dimensional QKD with the time-frequency entangled states \cite{Nunn2013, Mower2013}. The time-frequency encoding provides numerous advantages over other coding methods in the sense of dense multiplexing transmission in long-haul single-mode fibers.

\textbf{\emph{Acknowledgements}}
The authors are grateful to I. A. Walmsley and C. Ren for insightful discussions, and to K. Wakui for assistance in experiment. This work is supported by MEXT Grant-in-Aid for Young Scientists (B)15K17477, and the ImPACT Program of Council for Science, Technology and Innovation (Cabinet Office, Government of Japan).

\onecolumngrid

\section{Supplementary Information}
\subsection{S1. Frequency entangled qudits}
\begin{figure}[bp]
\includegraphics[width=0.38 \textwidth]{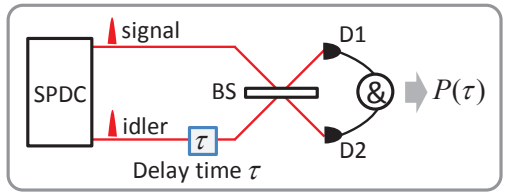}
\caption{The setup of Hong-Ou-Mandel (HOM) interference.}
\label{Fig1}
\end{figure}

Here we deduce the equations for the frequency-entangled qudit state generated in a Hong-Ou-Mandel (HOM) interference.
The setup of the scheme is shown in Fig.\,\ref{Fig1}.
The two-photon state from a spontaneous parametric down-conversion (SPDC) process can be described as
\begin{equation}\label{eq1}
\left| \Psi  \right\rangle  = \int_0^\infty  {\int_0^\infty  {d\omega _s d\omega _i } } f(\omega _s ,\omega _i )\hat a_s^\dag  (\omega _s )\hat a_i^\dag  (\omega _i )\left| {0} \right\rangle,
\end{equation}
where $\hat{a}^\dagger(\omega)$ is the creation operator at angular frequency
$\omega$, the subscripts $s$ and $i$ denote the signal and idler photons, respectively, and $f(\omega_s, \omega_i)$ is their joint spectral amplitude.

After a delay time $\tau$ on the idler arm and passing through a beamsplitter (BS), the state is evolved to be
\begin{equation}\label{eqs2}
\frac{{\rm{1}}}{{\rm{2}}} \int_0^\infty  {\int_0^\infty  {d\omega _s d\omega _i } }  f(\omega _s ,\omega _i ) e^{ - i\omega _i \tau }  [\hat a_1^\dag  (\omega _s )\hat a_1^\dag  (\omega _i ) - \hat a_2^\dag  (\omega _s )\hat a_2^\dag  (\omega _i ) - \hat a_1^\dag  (\omega _s )\hat a_2^\dag  (\omega _i ) + \hat a_2^\dag  (\omega _s )\hat a_1^\dag  (\omega _i )]   \left| {0} \right\rangle,
\end{equation}
In the above calculation, we applied the relation of
\begin{equation}\label{eqs3}
\begin{array}{l}
 \hat a_s^\dag  (\omega _s ) = \frac{{\rm{1}}}{{\sqrt {\rm{2}} }}[\hat a_1^\dag  (\omega _s ) + \hat a_2^\dag  (\omega _s )], \\ \\
 \hat a_i^\dag  (\omega _i )= \frac{{\rm{1}}}{{\sqrt {\rm{2}} }}[\hat a_1^\dag  (\omega _i ) - \hat a_2^\dag  (\omega _i )]e^{ - i\omega _i \tau },  \\
 \end{array}
\end{equation}
where $\hat a_1^\dag$  and $\hat a_2^\dag$  are the creation operators for the fields after the BS in channel 1 (connecting to detector D1) and channel 2 (connecting to detector  D2) respectively.
For the postselected state such that coincidence count occurs, only last two terms of Eq.\,(\ref{eqs2}) remains.
One is
\begin{equation}\label{eqs4}
 - \frac{{\rm{1}}}{{\rm{2}}}  \int_0^\infty  {\int_0^\infty  {d\omega _s d\omega _i } }  f(\omega _s ,\omega _i )e^{ - i\omega _i \tau } \hat a_1^\dag  (\omega _s )\hat a_2^\dag  (\omega _i )\left| {0} \right\rangle.
\end{equation}
The other is
\begin{equation}\label{eqs5}
  \frac{{\rm{1}}}{{\rm{2}}}  \int_0^\infty  {\int_0^\infty  {d\omega _s d\omega _i } }      f(\omega _s ,\omega _i )e^{ - i\omega _i \tau } \hat a_2^\dag  (\omega _s )\hat a_1^\dag  (\omega _i )\left| {0} \right\rangle  =  \frac{{\rm{1}}}{{\rm{2}}}   \int_0^\infty  {\int_0^\infty  {d\omega _s d\omega _i } }  f(\omega _i ,\omega _s )e^{ - i\omega _s \tau } \hat a_1^\dag  (\omega _s )\hat a_2^\dag  (\omega _i )\left| {0} \right\rangle .
\end{equation}
By combing these two terms, the unnormalized state after the BS is given by
\begin{equation}\label{eqs6}
  \int_0^\infty  {\int_0^\infty  {d\omega _s d\omega _i } }     [ - f(\omega _s ,\omega _i )e^{ - i\omega _i \tau }  + f(\omega _i ,\omega _s )e^{ - i\omega _s \tau } ]\hat a_1^\dag  (\omega _s )\hat a_2^\dag  (\omega _i )\left| {0} \right\rangle.
\end{equation}
Since we label the modes after the beamsplitter as  channel 1 and channel 2,
$\omega_s$ and $\omega_i$ should be replaced by $\omega_1$ and $\omega_2$,
respectively. Then the state (unnormalized) is
\begin{equation}\label{eqs7}
 \int_0^\infty  {\int_0^\infty  {d\omega _1 d\omega _2 } }    [ - f(\omega _1 ,\omega _2 )e^{ - i\omega _2 \tau }  + f(\omega _2 ,\omega _1 )e^{ - i\omega _1 \tau } ]\hat a_1^\dag  (\omega _1 )\hat a_2^\dag  (\omega _2 )\left| {0} \right\rangle.
\end{equation}
Cancelling the global phase by applying $ - e^{ i\omega _2 \tau }$ and
normalizing, we obtain
\begin{equation}\label{eqs8}
\left| \phi (\tau)  \right\rangle =
\frac{1}{\sqrt{\mathcal{N}}}   \int_0^\infty  {\int_0^\infty  {d\omega _1 d\omega _2 } }      [f(\omega _1 ,\omega _2 ) - f(\omega _2 ,\omega _1 )e^{ - i(\omega _1  - \omega _2 )\tau } ]\hat a_1^\dag  (\omega _1 )\hat a_2^\dag  (\omega _2 )\left| {0} \right\rangle         ,
\end{equation}
where
\begin{eqnarray}\label{norm}
\mathcal{N} & = & \int_0^\infty \int_0^\infty d \omega_1 d \omega_2
\left| f(\omega _1 ,\omega _2 ) - f(\omega _2 ,\omega _1 )e^{ - i(\omega _1  - \omega _2 )\tau } \right|^2
\nonumber\\ & = &
\int_0^\infty \int_0^\infty d \omega_1 d \omega_2
I(\omega_1, \omega_2, \tau) ,
\end{eqnarray}
is the normalization factor.

Due to the strongly anti-correlated phase matching of the PPSLT crystal,
$f(\omega_1, \omega_2)$ is approximately described by
\begin{equation}\label{JSA_approx}
f(\omega_s, \omega_i) \approx
h(\omega_s + \omega_i - \omega_p) \delta (\omega_s + \omega_i - \omega_p).
\end{equation}
where $\omega_p$ is the pump frequency,
$\delta(x)$ is the Dirac delta function, and
$h(x)$ is determined by the phase matching condition and the pump envelope.
Substituting this into Eq.~(\ref{eqs8}), we have
\begin{equation}\label{EntQudit}
|\phi (\tau)\rangle \approx  \frac{1}{\sqrt{\mathcal{N}}}
\int^\infty_0 \int^\infty_0 d\omega_1 d\omega_2
h(\omega_1 + \omega_2 - \omega_p) \delta (\omega_1 + \omega_2 - \omega_p)
(1-e^{-i(\omega_1-\omega_2)\tau})
\hat{a}_1^\dagger (\omega_1) \hat{a}_2^\dagger (\omega_2) |0\rangle .
\end{equation}

\subsection{S2. The HOM detection probability
and the correlated spectral intensity (CSI)}
In the following, we describe the detection process
in our HOM experiment.
For simplicity, we denote
\begin{equation}\label{gfunc}
g(\omega_1, \omega_2, \tau) \equiv
f(\omega_1, \omega_2) - f(\omega_2, \omega_1) e^{-i(\omega_1-\omega_2)\tau}.
\end{equation}
The detection field operators of detector 1 (D1) and detector 2 (D2) are
\begin{equation}\label{eqs10}
\begin{array}{l}
 \hat E_1^{( + )} (t_1 ) = \frac{1}{{\sqrt {2\pi } }}\int_0^\infty  {d\omega _1 } \hat a_1 (\omega _1 )e^{ - i\omega _1 t_1 } , \\\\
 \hat E_2^{( + )} (t_2 ) = \frac{1}{{\sqrt {2\pi } }}\int_0^\infty  {d\omega _2 \hat a_2 (\omega _2 )} e^{ - i\omega _2 t_2 } . \\
 \end{array}
\end{equation}
The coincidence probability $P(\tau )$ as a function of delay time can be expressed as
\begin{equation}\label{eqs11}
P(\tau ) = \int {\int {dt_1 dt_2 } } \left\langle {\phi \left| {\hat E_1^{( - )} \hat E_2^{( - )} \hat E_2^{( + )} \hat E_1^{( + )} } \right|\phi } \right\rangle.
\end{equation}
First consider $\hat E_2^{( + )} \hat E_1^{( + )} \left| \phi  \right\rangle$,
\begin{equation}\label{eqs12}
\begin{array}{lll}
 \hat E_2^{( + )} \hat E_1^{( + )} \left| \phi  \right\rangle
 &=& \frac{1}{{2\pi }} \int_0^\infty \int_0^\infty  {d\omega _1 }  {d\omega _2 \hat a_1 (\omega _1 )\hat a_2 (\omega _2 )e^{ - i\omega _1 t_1 } } e^{ - i\omega _2 t_2 }  \times \frac{1}{2}\int_0^\infty \int_0^\infty {d\omega _1^,   {d\omega _2^, } } g(\omega _1^, ,\omega _2^, ,\tau )\hat a_1^\dag  (\omega _1^, )\hat a_2^\dag  (\omega _2^, )\left| {0} \right\rangle  \\ \\
  &=& \frac{1}{{4\pi }}\int_0^\infty \int_0^\infty \int_0^\infty \int_0^\infty  {d\omega _1 }  {d\omega _2 }  {d\omega _1^,  {d\omega _2^, } } g(\omega _1^, ,\omega _2^, ,\tau )e^{ - i\omega _1 t_1 } e^{ - i\omega _2 t_2 } \delta (\omega _1  - \omega _1^, )\delta (\omega _2  - \omega _2^, )\left| {0} \right\rangle  \\ \\
  &=& \frac{1}{{4\pi }}\int_0^\infty \int_0^\infty  {d\omega _1 }  {d\omega _2 g(\omega _1 ,\omega _2 ,\tau )e^{ - i\omega _1 t_1 } } e^{ - i\omega _2 t_2 } \left| {0} \right\rangle . \\
 \end{array}
\end{equation}
In the above calculation,
$\hat{a}(\omega)\hat{a}^\dagger(\omega^,) - \hat{a}^\dagger(\omega)\hat{a}(\omega^,)= \delta (\omega  - \omega ^, )$ and $\hat{a}(\omega)|0\rangle=0$ are used.
Then,
\begin{equation}\label{eqs13}
\begin{array}{l}
\left\langle {\phi \left| {\hat E_1^{( - )} \hat E_2^{( - )} \hat E_2^{( + )} \hat E_1^{( + )} } \right|\phi } \right\rangle \\\\
 = (\frac{1}{{4\pi }})^2 \int_0^\infty \int_0^\infty  {d\omega _1 } {d\omega _2 g(\omega _1 ,\omega _2 ,\tau )e^{ - i\omega _1 t_1 } } e^{ - i\omega _2 t_2 }  \times \int_0^\infty \int_0^\infty  {d\omega _1^, }  {d\omega _2^, g^* (\omega _1^, ,\omega _2^, ,\tau )e^{  i\omega _1^, t_1 } } e^{  i\omega _2^, t_2 }.\\
\end{array}
\end{equation}
Finally,
\begin{equation}\label{eqs14}
\begin{array}{lll}
 P(\tau ) &=& \int {\int {dt_1 dt_2 } } \left\langle {\phi \left| {\hat E_1^{( - )} \hat E_2^{( - )} \hat E_2^{( + )} \hat E_1^{( + )} } \right|\phi } \right\rangle  \\\\
  &=& (\frac{1}{{4\pi }})^2 \int {dt_1 dt_2 } d\omega _1 d\omega _2 d\omega _1^, d\omega _2^, g(\omega _1 ,\omega _2 ,\tau )g^* (\omega _1^, ,\omega _2^, ,\tau ) e^{ - i(\omega _1  - \omega _1^, )t_1 } e^{ - i(\omega _2  - \omega _2^, )t_2 } \\\\
  &=& \frac{1}{4} \int_0^\infty  \int_0^\infty  \int_0^\infty \int_0^\infty  {d\omega _1   {d\omega _2   {d\omega _1^,  {d\omega _2^, } } } }  g(\omega _1 ,\omega _2 ,\tau )g^* (\omega _1^, ,\omega _2^, ,\tau )   \delta (\omega _1  - \omega _1^, )\delta (\omega _2  - \omega _2^, ) \\\\
  &=& \frac{1}{4}\int_0^\infty \int_0^\infty  {d\omega _1 }   {d\omega _2 } {\rm{|}}g(\omega _1 ,\omega _2 ,\tau ){\rm{|}}^{\rm{2}} . \\
 \end{array}
\end{equation}
In the above calculation, the relation of $\delta (\omega  - \omega ^, ) = \frac{1}{{2\pi }}\int_{ - \infty }^\infty  {e^{-i(\omega  - \omega ^, )t} } dt$ is used.
If $f(\omega _1 ,\omega _2 )$ is real, i.e. $f(\omega _1 ,\omega _2 )=f^*(\omega _1 ,\omega _2 )$, the correlated spectral intensity (CSI) can be further simplified as
\begin{equation}\label{eqs15}
I(\omega _1 ,\omega _2 ,\tau ){\rm{  \equiv  |}}g(\omega _1 ,\omega _2 ,\tau ){\rm{|}}^{\rm{2}}  = {\rm{|}}f(\omega _1 ,\omega _2 ){\rm{|}}^{\rm{2}}  + {\rm{|}}f(\omega _2 ,\omega _1 ){\rm{|}}^{\rm{2}}  - 2f(\omega _1 ,\omega _2 )f(\omega _2 ,\omega _1 )cos(\omega _1  - \omega _2 )\tau.
\end{equation}


\begin{thebibliography}{34}
\expandafter\ifx\csname natexlab\endcsname\relax\def\natexlab#1{#1}\fi
\expandafter\ifx\csname bibnamefont\endcsname\relax
  \def\bibnamefont#1{#1}\fi
\expandafter\ifx\csname bibfnamefont\endcsname\relax
  \def\bibfnamefont#1{#1}\fi
\expandafter\ifx\csname citenamefont\endcsname\relax
  \def\citenamefont#1{#1}\fi
\expandafter\ifx\csname url\endcsname\relax
  \def\url#1{\texttt{#1}}\fi
\expandafter\ifx\csname urlprefix\endcsname\relax\def\urlprefix{URL }\fi
\providecommand{\bibinfo}[2]{#2}
\providecommand{\eprint}[2][]{\url{#2}}

\bibitem[{\citenamefont{Kaszlikowski et~al.}(2000)\citenamefont{Kaszlikowski,
  Gnaci{\'n}ski, { \.Z}ukowski, Miklaszewski, and
  Zeilinger}}]{Kaszlikowski2000}
\bibinfo{author}{\bibfnamefont{D.}~\bibnamefont{Kaszlikowski}},
  \bibinfo{author}{\bibfnamefont{P.}~\bibnamefont{Gnaci{\'n}ski}},
  \bibinfo{author}{\bibfnamefont{M.}~\bibnamefont{{ \.Z}ukowski}},
  \bibinfo{author}{\bibfnamefont{W.}~\bibnamefont{Miklaszewski}},
  \bibnamefont{and}
  \bibinfo{author}{\bibfnamefont{A.}~\bibnamefont{Zeilinger}},
  \bibinfo{journal}{Phys. Rev. Lett.} \textbf{\bibinfo{volume}{85}},
  \bibinfo{pages}{4418} (\bibinfo{year}{2000}).

\bibitem[{\citenamefont{Lapkiewicz et~al.}(2011)\citenamefont{Lapkiewicz, Li,
  Schaeff, Langford, Ramelow, Wiesniak, and Zeilinger}}]{Lapkiewicz2011}
\bibinfo{author}{\bibfnamefont{R.}~\bibnamefont{Lapkiewicz}},
  \bibinfo{author}{\bibfnamefont{P.}~\bibnamefont{Li}},
  \bibinfo{author}{\bibfnamefont{C.}~\bibnamefont{Schaeff}},
  \bibinfo{author}{\bibfnamefont{N.~K.} \bibnamefont{Langford}},
  \bibinfo{author}{\bibfnamefont{S.}~\bibnamefont{Ramelow}},
  \bibinfo{author}{\bibfnamefont{M.}~\bibnamefont{Wiesniak}}, \bibnamefont{and}
  \bibinfo{author}{\bibfnamefont{A.}~\bibnamefont{Zeilinger}},
  \bibinfo{journal}{Nature} \textbf{\bibinfo{volume}{474}},
  \bibinfo{pages}{490} (\bibinfo{year}{2011}).

\bibitem[{\citenamefont{Durt et~al.}(2004)\citenamefont{Durt, Kaszlikowski,
  Chen, and Kwek}}]{Durt2004}
\bibinfo{author}{\bibfnamefont{T.}~\bibnamefont{Durt}},
  \bibinfo{author}{\bibfnamefont{D.}~\bibnamefont{Kaszlikowski}},
  \bibinfo{author}{\bibfnamefont{J.-L.} \bibnamefont{Chen}}, \bibnamefont{and}
  \bibinfo{author}{\bibfnamefont{L.~C.} \bibnamefont{Kwek}},
  \bibinfo{journal}{Phys. Rev. A} \textbf{\bibinfo{volume}{69}},
  \bibinfo{pages}{032313} (\bibinfo{year}{2004}).

\bibitem[{\citenamefont{Nunn et~al.}(2013)\citenamefont{Nunn, Wright,
  S\"{o}ller, Zhang, Walmsley, and Smith}}]{Nunn2013}
\bibinfo{author}{\bibfnamefont{J.}~\bibnamefont{Nunn}},
  \bibinfo{author}{\bibfnamefont{L.~J.} \bibnamefont{Wright}},
  \bibinfo{author}{\bibfnamefont{C.}~\bibnamefont{S\"{o}ller}},
  \bibinfo{author}{\bibfnamefont{L.}~\bibnamefont{Zhang}},
  \bibinfo{author}{\bibfnamefont{I.~A.} \bibnamefont{Walmsley}},
  \bibnamefont{and} \bibinfo{author}{\bibfnamefont{B.~J.} \bibnamefont{Smith}},
  \bibinfo{journal}{Opt. Express} \textbf{\bibinfo{volume}{21}},
  \bibinfo{pages}{15959} (\bibinfo{year}{2013}).

\bibitem[{\citenamefont{Mower et~al.}(2013)\citenamefont{Mower, Zhang,
  Desjardins, Lee, Shapiro, and Englund}}]{Mower2013}
\bibinfo{author}{\bibfnamefont{J.}~\bibnamefont{Mower}},
  \bibinfo{author}{\bibfnamefont{Z.}~\bibnamefont{Zhang}},
  \bibinfo{author}{\bibfnamefont{P.}~\bibnamefont{Desjardins}},
  \bibinfo{author}{\bibfnamefont{C.}~\bibnamefont{Lee}},
  \bibinfo{author}{\bibfnamefont{J.~H.} \bibnamefont{Shapiro}},
  \bibnamefont{and} \bibinfo{author}{\bibfnamefont{D.}~\bibnamefont{Englund}},
  \bibinfo{journal}{Phys. Rev. A} \textbf{\bibinfo{volume}{87}},
  \bibinfo{pages}{062322} (\bibinfo{year}{2013}).

\bibitem[{\citenamefont{Langford et~al.}(2004)\citenamefont{Langford, Dalton,
  Harvey, O'Brien, Pryde, Gilchrist, Bartlett, and White}}]{Langford2004}
\bibinfo{author}{\bibfnamefont{N.~K.} \bibnamefont{Langford}},
  \bibinfo{author}{\bibfnamefont{R.~B.} \bibnamefont{Dalton}},
  \bibinfo{author}{\bibfnamefont{M.~D.} \bibnamefont{Harvey}},
  \bibinfo{author}{\bibfnamefont{J.~L.} \bibnamefont{O'Brien}},
  \bibinfo{author}{\bibfnamefont{G.~J.} \bibnamefont{Pryde}},
  \bibinfo{author}{\bibfnamefont{A.}~\bibnamefont{Gilchrist}},
  \bibinfo{author}{\bibfnamefont{S.~D.} \bibnamefont{Bartlett}},
  \bibnamefont{and} \bibinfo{author}{\bibfnamefont{A.~G.} \bibnamefont{White}},
  \bibinfo{journal}{Phys. Rev. Lett.} \textbf{\bibinfo{volume}{93}},
  \bibinfo{pages}{053601} (\bibinfo{year}{2004}).

\bibitem[{\citenamefont{Howland and Howell}(2013)}]{Howland2013}
\bibinfo{author}{\bibfnamefont{G.~A.} \bibnamefont{Howland}} \bibnamefont{and}
  \bibinfo{author}{\bibfnamefont{J.~C.} \bibnamefont{Howell}},
  \bibinfo{journal}{Phys. Rev. X} \textbf{\bibinfo{volume}{3}},
  \bibinfo{pages}{011013} (\bibinfo{year}{2013}).

\bibitem[{\citenamefont{Bogdanov et~al.}(2004)\citenamefont{Bogdanov, Chekhova,
  Kulik, Maslennikov, Zhukov., Oh, and Tey}}]{Bogdanov2004}
\bibinfo{author}{\bibfnamefont{Y.~I.} \bibnamefont{Bogdanov}},
  \bibinfo{author}{\bibfnamefont{M.~V.} \bibnamefont{Chekhova}},
  \bibinfo{author}{\bibfnamefont{S.~P.} \bibnamefont{Kulik}},
  \bibinfo{author}{\bibfnamefont{G.~A.} \bibnamefont{Maslennikov}},
  \bibinfo{author}{\bibfnamefont{A.~A.} \bibnamefont{Zhukov.}},
  \bibinfo{author}{\bibfnamefont{C.~H.} \bibnamefont{Oh}}, \bibnamefont{and}
  \bibinfo{author}{\bibfnamefont{M.~K.} \bibnamefont{Tey}},
  \bibinfo{journal}{Phys. Rev. Lett.} \textbf{\bibinfo{volume}{93}},
  \bibinfo{pages}{230503} (\bibinfo{year}{2004}).

\bibitem[{\citenamefont{Bernhard et~al.}(2013)\citenamefont{Bernhard, Bessire,
  Feurer, and Stefanov}}]{Bernhard2013}
\bibinfo{author}{\bibfnamefont{C.}~\bibnamefont{Bernhard}},
  \bibinfo{author}{\bibfnamefont{B.}~\bibnamefont{Bessire}},
  \bibinfo{author}{\bibfnamefont{T.}~\bibnamefont{Feurer}}, \bibnamefont{and}
  \bibinfo{author}{\bibfnamefont{A.}~\bibnamefont{Stefanov}},
  \bibinfo{journal}{Phys. Rev. A} \textbf{\bibinfo{volume}{88}},
  \bibinfo{pages}{032322} (\bibinfo{year}{2013}).

\bibitem[{\citenamefont{Thew et~al.}(2004)\citenamefont{Thew, Ac{\'i}n,
  Zbinden, and Gisin}}]{Thew2004}
\bibinfo{author}{\bibfnamefont{R.~T.} \bibnamefont{Thew}},
  \bibinfo{author}{\bibfnamefont{A.}~\bibnamefont{Ac{\'i}n}},
  \bibinfo{author}{\bibfnamefont{H.}~\bibnamefont{Zbinden}}, \bibnamefont{and}
  \bibinfo{author}{\bibfnamefont{N.}~\bibnamefont{Gisin}},
  \bibinfo{journal}{Phys. Rev. Lett.} \textbf{\bibinfo{volume}{93}},
  \bibinfo{pages}{010503} (\bibinfo{year}{2004}).

\bibitem[{\citenamefont{Bessire et~al.}(2014)\citenamefont{Bessire, Bernhard,
  Feurer, and Stefanov}}]{Bessire2014}
\bibinfo{author}{\bibfnamefont{B.}~\bibnamefont{Bessire}},
  \bibinfo{author}{\bibfnamefont{C.}~\bibnamefont{Bernhard}},
  \bibinfo{author}{\bibfnamefont{T.}~\bibnamefont{Feurer}}, \bibnamefont{and}
  \bibinfo{author}{\bibfnamefont{A.}~\bibnamefont{Stefanov}},
  \bibinfo{journal}{New J. Phys.} \textbf{\bibinfo{volume}{16}},
  \bibinfo{pages}{033017} (\bibinfo{year}{2014}).

\bibitem[{\citenamefont{Krenn et~al.}(2014)\citenamefont{Krenn, Huber, Fickler,
  Lapkiewicz, Ramelow, and Zeilinger}}]{Krenn2014}
\bibinfo{author}{\bibfnamefont{M.}~\bibnamefont{Krenn}},
  \bibinfo{author}{\bibfnamefont{M.}~\bibnamefont{Huber}},
  \bibinfo{author}{\bibfnamefont{R.}~\bibnamefont{Fickler}},
  \bibinfo{author}{\bibfnamefont{R.}~\bibnamefont{Lapkiewicz}},
  \bibinfo{author}{\bibfnamefont{S.}~\bibnamefont{Ramelow}}, \bibnamefont{and}
  \bibinfo{author}{\bibfnamefont{A.}~\bibnamefont{Zeilinger}},
  \bibinfo{journal}{PNAS} \textbf{\bibinfo{volume}{111}}, \bibinfo{pages}{6243}
  (\bibinfo{year}{2014}).

\bibitem[{\citenamefont{Olislager et~al.}(2010)\citenamefont{Olislager, Cussey,
  Nguyen, Emplit, Massar, Merolla, and Huy}}]{Olislager2010}
\bibinfo{author}{\bibfnamefont{L.}~\bibnamefont{Olislager}},
  \bibinfo{author}{\bibfnamefont{J.}~\bibnamefont{Cussey}},
  \bibinfo{author}{\bibfnamefont{A.~T.} \bibnamefont{Nguyen}},
  \bibinfo{author}{\bibfnamefont{P.}~\bibnamefont{Emplit}},
  \bibinfo{author}{\bibfnamefont{S.}~\bibnamefont{Massar}},
  \bibinfo{author}{\bibfnamefont{J.-M.} \bibnamefont{Merolla}},
  \bibnamefont{and} \bibinfo{author}{\bibfnamefont{K.~P.} \bibnamefont{Huy}},
  \bibinfo{journal}{Phys. Rev. A} \textbf{\bibinfo{volume}{82}},
  \bibinfo{pages}{013804} (\bibinfo{year}{2010}).

\bibitem[{\citenamefont{Shimizu and Edamatsu}(2009)}]{Shimizu2009}
\bibinfo{author}{\bibfnamefont{R.}~\bibnamefont{Shimizu}} \bibnamefont{and}
  \bibinfo{author}{\bibfnamefont{K.}~\bibnamefont{Edamatsu}},
  \bibinfo{journal}{Opt. Express} \textbf{\bibinfo{volume}{17}},
  \bibinfo{pages}{16385} (\bibinfo{year}{2009}).

\bibitem[{\citenamefont{Bisht and Shimizu}(2015)}]{Bisht2015}
\bibinfo{author}{\bibfnamefont{N.~S.} \bibnamefont{Bisht}} \bibnamefont{and}
  \bibinfo{author}{\bibfnamefont{R.}~\bibnamefont{Shimizu}},
  \bibinfo{journal}{J. Opt. Soc. Am. B} \textbf{\bibinfo{volume}{32}},
  \bibinfo{pages}{550} (\bibinfo{year}{2015}).

\bibitem[{\citenamefont{Edamatsu et~al.}(2011)\citenamefont{Edamatsu, Shimizu,
  Ueno, Jin, Kaneda, Yabuno, Suzuki, Nagano, Syouji, and Suizu}}]{Edamatsu2011}
\bibinfo{author}{\bibfnamefont{K.}~\bibnamefont{Edamatsu}},
  \bibinfo{author}{\bibfnamefont{R.}~\bibnamefont{Shimizu}},
  \bibinfo{author}{\bibfnamefont{W.}~\bibnamefont{Ueno}},
  \bibinfo{author}{\bibfnamefont{R.-B.} \bibnamefont{Jin}},
  \bibinfo{author}{\bibfnamefont{F.}~\bibnamefont{Kaneda}},
  \bibinfo{author}{\bibfnamefont{M.}~\bibnamefont{Yabuno}},
  \bibinfo{author}{\bibfnamefont{H.}~\bibnamefont{Suzuki}},
  \bibinfo{author}{\bibfnamefont{S.}~\bibnamefont{Nagano}},
  \bibinfo{author}{\bibfnamefont{A.}~\bibnamefont{Syouji}}, \bibnamefont{and}
  \bibinfo{author}{\bibfnamefont{K.}~\bibnamefont{Suizu}},
  \bibinfo{journal}{Prog. Inform.} \textbf{\bibinfo{volume}{8}},
  \bibinfo{pages}{19} (\bibinfo{year}{2011}).

\bibitem[{\citenamefont{Hong et~al.}(1987)\citenamefont{Hong, Ou, and
  Mandel}}]{Hong1987}
\bibinfo{author}{\bibfnamefont{C.~K.} \bibnamefont{Hong}},
  \bibinfo{author}{\bibfnamefont{Z.~Y.} \bibnamefont{Ou}}, \bibnamefont{and}
  \bibinfo{author}{\bibfnamefont{L.}~\bibnamefont{Mandel}},
  \bibinfo{journal}{Phys. Rev. Lett.} \textbf{\bibinfo{volume}{59}},
  \bibinfo{pages}{2044} (\bibinfo{year}{1987}).

\bibitem[{\citenamefont{Gerrits et~al.}(2015)\citenamefont{Gerrits, Marsili,
  Verma, Shalm, Shaw, Mirin, and Nam}}]{Gerrits2015}
\bibinfo{author}{\bibfnamefont{T.}~\bibnamefont{Gerrits}},
  \bibinfo{author}{\bibfnamefont{F.}~\bibnamefont{Marsili}},
  \bibinfo{author}{\bibfnamefont{V.~B.} \bibnamefont{Verma}},
  \bibinfo{author}{\bibfnamefont{L.~K.} \bibnamefont{Shalm}},
  \bibinfo{author}{\bibfnamefont{M.}~\bibnamefont{Shaw}},
  \bibinfo{author}{\bibfnamefont{R.~P.} \bibnamefont{Mirin}}, \bibnamefont{and}
  \bibinfo{author}{\bibfnamefont{S.~W.} \bibnamefont{Nam}},
  \bibinfo{journal}{Phys. Rev. A} \textbf{\bibinfo{volume}{91}},
  \bibinfo{pages}{013830} (\bibinfo{year}{2015}).

\bibitem[{\citenamefont{Jin et~al.}(2015{\natexlab{a}})\citenamefont{Jin,
  Gerrits, Fujiwara, Wakabayashi, Yamashita, Miki, Terai, Shimizu, Takeoka, and
  Sasaki}}]{Jin2015OE}
\bibinfo{author}{\bibfnamefont{R.-B.} \bibnamefont{Jin}},
  \bibinfo{author}{\bibfnamefont{T.}~\bibnamefont{Gerrits}},
  \bibinfo{author}{\bibfnamefont{M.}~\bibnamefont{Fujiwara}},
  \bibinfo{author}{\bibfnamefont{R.}~\bibnamefont{Wakabayashi}},
  \bibinfo{author}{\bibfnamefont{T.}~\bibnamefont{Yamashita}},
  \bibinfo{author}{\bibfnamefont{S.}~\bibnamefont{Miki}},
  \bibinfo{author}{\bibfnamefont{H.}~\bibnamefont{Terai}},
  \bibinfo{author}{\bibfnamefont{R.}~\bibnamefont{Shimizu}},
  \bibinfo{author}{\bibfnamefont{M.}~\bibnamefont{Takeoka}}, \bibnamefont{and}
  \bibinfo{author}{\bibfnamefont{M.}~\bibnamefont{Sasaki}},
  \bibinfo{journal}{Opt. Express} \textbf{\bibinfo{volume}{23}},
  \bibinfo{pages}{28836} (\bibinfo{year}{2015}{\natexlab{a}}).

\bibitem[{\citenamefont{U'Ren et~al.}(2006)\citenamefont{U'Ren, Erdmann, de~la
  Cruz-Gutierrez, and Walmsley}}]{URen2006}
\bibinfo{author}{\bibfnamefont{A.~B.} \bibnamefont{U'Ren}},
  \bibinfo{author}{\bibfnamefont{R.~K.} \bibnamefont{Erdmann}},
  \bibinfo{author}{\bibfnamefont{M.}~\bibnamefont{de~la Cruz-Gutierrez}},
  \bibnamefont{and} \bibinfo{author}{\bibfnamefont{I.~A.}
  \bibnamefont{Walmsley}}, \bibinfo{journal}{Phys. Rev. Lett.}
  \textbf{\bibinfo{volume}{97}}, \bibinfo{pages}{223602}
  (\bibinfo{year}{2006}).

\bibitem[{\citenamefont{Jin et~al.}(2013)\citenamefont{Jin, Shimizu, Wakui,
  Benichi, and Sasaki}}]{Jin2013OE}
\bibinfo{author}{\bibfnamefont{R.-B.} \bibnamefont{Jin}},
  \bibinfo{author}{\bibfnamefont{R.}~\bibnamefont{Shimizu}},
  \bibinfo{author}{\bibfnamefont{K.}~\bibnamefont{Wakui}},
  \bibinfo{author}{\bibfnamefont{H.}~\bibnamefont{Benichi}}, \bibnamefont{and}
  \bibinfo{author}{\bibfnamefont{M.}~\bibnamefont{Sasaki}},
  \bibinfo{journal}{Opt. Express} \textbf{\bibinfo{volume}{21}},
  \bibinfo{pages}{10659} (\bibinfo{year}{2013}).

\bibitem[{\citenamefont{Grice and Walmsley}(1997)}]{Grice1997}
\bibinfo{author}{\bibfnamefont{W.~P.} \bibnamefont{Grice}} \bibnamefont{and}
  \bibinfo{author}{\bibfnamefont{I.~A.} \bibnamefont{Walmsley}},
  \bibinfo{journal}{Phys. Rev. A} \textbf{\bibinfo{volume}{56}},
  \bibinfo{pages}{1627} (\bibinfo{year}{1997}).

\bibitem[{\citenamefont{Avenhaus et~al.}(2009)\citenamefont{Avenhaus, Eckstein,
  Mosley, and Silberhorn}}]{Avenhaus2009}
\bibinfo{author}{\bibfnamefont{M.}~\bibnamefont{Avenhaus}},
  \bibinfo{author}{\bibfnamefont{A.}~\bibnamefont{Eckstein}},
  \bibinfo{author}{\bibfnamefont{P.~J.} \bibnamefont{Mosley}},
  \bibnamefont{and}
  \bibinfo{author}{\bibfnamefont{C.}~\bibnamefont{Silberhorn}},
  \bibinfo{journal}{Opt. Lett.} \textbf{\bibinfo{volume}{34}},
  \bibinfo{pages}{2873} (\bibinfo{year}{2009}).

\bibitem[{\citenamefont{Gerrits et~al.}(2011)\citenamefont{Gerrits, Stevens,
  Baek, Calkins, Lita, Glancy, Knill, Nam, Mirin, Hadfield
  et~al.}}]{Gerrits2011}
\bibinfo{author}{\bibfnamefont{T.}~\bibnamefont{Gerrits}},
  \bibinfo{author}{\bibfnamefont{M.~J.} \bibnamefont{Stevens}},
  \bibinfo{author}{\bibfnamefont{B.}~\bibnamefont{Baek}},
  \bibinfo{author}{\bibfnamefont{B.}~\bibnamefont{Calkins}},
  \bibinfo{author}{\bibfnamefont{A.}~\bibnamefont{Lita}},
  \bibinfo{author}{\bibfnamefont{S.}~\bibnamefont{Glancy}},
  \bibinfo{author}{\bibfnamefont{E.}~\bibnamefont{Knill}},
  \bibinfo{author}{\bibfnamefont{S.~W.} \bibnamefont{Nam}},
  \bibinfo{author}{\bibfnamefont{R.~P.} \bibnamefont{Mirin}},
  \bibinfo{author}{\bibfnamefont{R.~H.} \bibnamefont{Hadfield}},
  \bibnamefont{et~al.}, \bibinfo{journal}{Opt. Express}
  \textbf{\bibinfo{volume}{19}}, \bibinfo{pages}{24434} (\bibinfo{year}{2011}).

\bibitem[{\citenamefont{Miki et~al.}(2013)\citenamefont{Miki, Yamashita, Terai,
  and Wang}}]{Miki2013}
\bibinfo{author}{\bibfnamefont{S.}~\bibnamefont{Miki}},
  \bibinfo{author}{\bibfnamefont{T.}~\bibnamefont{Yamashita}},
  \bibinfo{author}{\bibfnamefont{H.}~\bibnamefont{Terai}}, \bibnamefont{and}
  \bibinfo{author}{\bibfnamefont{Z.}~\bibnamefont{Wang}},
  \bibinfo{journal}{Opt. Express} \textbf{\bibinfo{volume}{21}},
  \bibinfo{pages}{10208} (\bibinfo{year}{2013}).

\bibitem[{\citenamefont{Yamashita et~al.}(2013)\citenamefont{Yamashita, Miki,
  Terai, and Wang}}]{Yamashita2013}
\bibinfo{author}{\bibfnamefont{T.}~\bibnamefont{Yamashita}},
  \bibinfo{author}{\bibfnamefont{S.}~\bibnamefont{Miki}},
  \bibinfo{author}{\bibfnamefont{H.}~\bibnamefont{Terai}}, \bibnamefont{and}
  \bibinfo{author}{\bibfnamefont{Z.}~\bibnamefont{Wang}},
  \bibinfo{journal}{Opt. Express} \textbf{\bibinfo{volume}{21}},
  \bibinfo{pages}{27177} (\bibinfo{year}{2013}).

\bibitem[{\citenamefont{Jin et~al.}(2015{\natexlab{b}})\citenamefont{Jin,
  Fujiwara, Yamashita, Miki, Terai, Wang, Wakui, Shimizu, and
  Sasaki}}]{Jin2015OC}
\bibinfo{author}{\bibfnamefont{R.-B.} \bibnamefont{Jin}},
  \bibinfo{author}{\bibfnamefont{M.}~\bibnamefont{Fujiwara}},
  \bibinfo{author}{\bibfnamefont{T.}~\bibnamefont{Yamashita}},
  \bibinfo{author}{\bibfnamefont{S.}~\bibnamefont{Miki}},
  \bibinfo{author}{\bibfnamefont{H.}~\bibnamefont{Terai}},
  \bibinfo{author}{\bibfnamefont{Z.}~\bibnamefont{Wang}},
  \bibinfo{author}{\bibfnamefont{K.}~\bibnamefont{Wakui}},
  \bibinfo{author}{\bibfnamefont{R.}~\bibnamefont{Shimizu}}, \bibnamefont{and}
  \bibinfo{author}{\bibfnamefont{M.}~\bibnamefont{Sasaki}},
  \bibinfo{journal}{Opt. Commun.} \textbf{\bibinfo{volume}{336}},
  \bibinfo{pages}{47} (\bibinfo{year}{2015}{\natexlab{b}}).

\bibitem[{\citenamefont{Tanaka et~al.}(2012)\citenamefont{Tanaka, Okamoto, Lim,
  Subashchandran, Okano, Zhang, Kang, Chen, Wu, Hirohata et~al.}}]{Tanaka2012}
\bibinfo{author}{\bibfnamefont{A.}~\bibnamefont{Tanaka}},
  \bibinfo{author}{\bibfnamefont{R.}~\bibnamefont{Okamoto}},
  \bibinfo{author}{\bibfnamefont{H.~H.} \bibnamefont{Lim}},
  \bibinfo{author}{\bibfnamefont{S.}~\bibnamefont{Subashchandran}},
  \bibinfo{author}{\bibfnamefont{M.}~\bibnamefont{Okano}},
  \bibinfo{author}{\bibfnamefont{L.}~\bibnamefont{Zhang}},
  \bibinfo{author}{\bibfnamefont{L.}~\bibnamefont{Kang}},
  \bibinfo{author}{\bibfnamefont{J.}~\bibnamefont{Chen}},
  \bibinfo{author}{\bibfnamefont{P.}~\bibnamefont{Wu}},
  \bibinfo{author}{\bibfnamefont{T.}~\bibnamefont{Hirohata}},
  \bibnamefont{et~al.}, \bibinfo{journal}{Opt. Express}
  \textbf{\bibinfo{volume}{20}}, \bibinfo{pages}{25228} (\bibinfo{year}{2012}).

\bibitem[{\citenamefont{James et~al.}(2001)\citenamefont{James, Kwiat, Munro,
  and White}}]{James2001}
\bibinfo{author}{\bibfnamefont{D.~F.~V.} \bibnamefont{James}},
  \bibinfo{author}{\bibfnamefont{P.~G.} \bibnamefont{Kwiat}},
  \bibinfo{author}{\bibfnamefont{W.~J.} \bibnamefont{Munro}}, \bibnamefont{and}
  \bibinfo{author}{\bibfnamefont{A.~G.} \bibnamefont{White}},
  \bibinfo{journal}{Phys. Rev. A} \textbf{\bibinfo{volume}{64}},
  \bibinfo{pages}{052312} (\bibinfo{year}{2001}).

\bibitem[{\citenamefont{Giovannetti et~al.}(2001)\citenamefont{Giovannetti,
  Lloyd, Maccone, and Wong}}]{Giovannetti2001}
\bibinfo{author}{\bibfnamefont{V.}~\bibnamefont{Giovannetti}},
  \bibinfo{author}{\bibfnamefont{S.}~\bibnamefont{Lloyd}},
  \bibinfo{author}{\bibfnamefont{L.}~\bibnamefont{Maccone}}, \bibnamefont{and}
  \bibinfo{author}{\bibfnamefont{F.~N.~C.} \bibnamefont{Wong}},
  \bibinfo{journal}{Phys. Rev. Lett.} \textbf{\bibinfo{volume}{87}},
  \bibinfo{pages}{117902} (\bibinfo{year}{2001}).

\bibitem[{\citenamefont{Ho et~al.}(2009)\citenamefont{Ho, Lamas-Linares, and
  Kurtsiefer}}]{Ho2009}
\bibinfo{author}{\bibfnamefont{C.}~\bibnamefont{Ho}},
  \bibinfo{author}{\bibfnamefont{A.}~\bibnamefont{Lamas-Linares}},
  \bibnamefont{and}
  \bibinfo{author}{\bibfnamefont{C.}~\bibnamefont{Kurtsiefer}},
  \bibinfo{journal}{New J. Phys.} \textbf{\bibinfo{volume}{11}},
  \bibinfo{pages}{045011} (\bibinfo{year}{2009}).

\bibitem[{\citenamefont{Quan et~al.}(2016)\citenamefont{Quan, Zhai, Wang, Hou,
  Xiang, Liu, Zhang, and Dong}}]{Quan2016}
\bibinfo{author}{\bibfnamefont{R.}~\bibnamefont{Quan}},
  \bibinfo{author}{\bibfnamefont{Y.}~\bibnamefont{Zhai}},
  \bibinfo{author}{\bibfnamefont{M.}~\bibnamefont{Wang}},
  \bibinfo{author}{\bibfnamefont{S.}~\bibnamefont{Hou},
  \bibfnamefont{Feiyanand~Wang}},
  \bibinfo{author}{\bibfnamefont{X.}~\bibnamefont{Xiang}},
  \bibinfo{author}{\bibfnamefont{T.}~\bibnamefont{Liu}},
  \bibinfo{author}{\bibfnamefont{S.}~\bibnamefont{Zhang}}, \bibnamefont{and}
  \bibinfo{author}{\bibfnamefont{R.}~\bibnamefont{Dong}},
  \bibinfo{journal}{arXiv:1602.06371}  (\bibinfo{year}{2016}).

\bibitem[{\citenamefont{Ren and Hofmann}(2012)}]{Ren2012}
\bibinfo{author}{\bibfnamefont{C.}~\bibnamefont{Ren}} \bibnamefont{and}
  \bibinfo{author}{\bibfnamefont{H.~F.} \bibnamefont{Hofmann}},
  \bibinfo{journal}{Phys. Rev. A} \textbf{\bibinfo{volume}{86}},
  \bibinfo{pages}{043823} (\bibinfo{year}{2012}).

\bibitem[{\citenamefont{Ren and Hofmann}(2014)}]{Ren2014}
\bibinfo{author}{\bibfnamefont{C.}~\bibnamefont{Ren}} \bibnamefont{and}
  \bibinfo{author}{\bibfnamefont{H.~F.} \bibnamefont{Hofmann}},
  \bibinfo{journal}{Phys. Rev. A} \textbf{\bibinfo{volume}{89}},
  \bibinfo{pages}{053823} (\bibinfo{year}{2014}).

\end{thebibliography}
\end{document}